\documentclass[aps,preprint,showpacs,groupedaddress]{revtex4}

\usepackage{graphicx}
\usepackage{dcolumn}
\usepackage{bm}

\begin{document}

\title{Tip-splitting evolution in the idealized Saffman-Taylor problem}

\author{Eldad Bettelheim}
\email[]{eldadb@phys.huji.ac.il}
\affiliation{James Franck Institute, University of Chicago, 5640
S. Ellis Ave., Chicago IL, 60637}
\author{Oded Agam}
\affiliation{Racah Institute of Physics, Hebrew University, Jerusalem 91904, Israel}

\date{\today}

\begin{abstract}
We derive a formula describing the evolution of tip-splittings of
Saffman-Taylor fingers in a Hele-Shaw cell, at zero surface tension.
\end{abstract}

\pacs{02.30.Ik, 05.45.Df, 05.45.Yv}

\maketitle
Fingered patterns characterize the non-equilibrium growth processes
of many systems, e.g.  the dendritic shape of snowflakes, the
outline of bacteria colonies growing in stressed environments, electro-deposition,
dielectric breakdown, and viscous fingering formed by forcing a non-viscous
fluid into the center of a Hele-Shaw cell filled with a viscous fluid
(Saffman-Taylor flow)\cite{Halsey:Today}. Repeated events of tip-splittings and side
branching during the growing process, manifest themselves in complicated
fractal patterns. Understanding the structure of such patterns is
a challenge for theorists.

When the growing patterns are self similar, one expects that
their global structure can be deduced from
the basic elements of the growing process, such as tip-splitting and
side-branching. Indeed, studies of theoretical models of diffusion
limited aggregation\cite{Witten:Sander}, and their generalizations
(e.g. the dielectric breakdown model) demonstrate that the fractal
dimension of the corresponding  patterns is related to characteristics
of the tip-splitting events\cite{170:Halsey} (side-branching in these models is negligible).

In this work we focus our attention on the fingered patterns
generated by Saffman-Taylor (ST) flows in a Hele-Shaw cell, in the
radial geometry.
 In particular, we shall
be interested in the evolution of tip-splitting in the singular
limit where the surface tension associated with the interface between
the viscous and the non-viscous fluids approaches zero (or alternatively
high pumping rate of the non-viscous fluid). The ST dynamics in this limit, which we shall refer
to as the idealized ST problem, has been shown to be integrable\cite{43:Zabr:Wiegm}.
Therefore, it constitutes an important paradigm in the field of non-equilibrium  growth processes.

The idealized ST problem admits a large class of exact solutions
\cite{186:Howison,R,185:Bensimon}. Among these one may find
solutions which resemble various forms of tips-split at large
surface tension\cite{HohlovHowison}. These solutions are
characterized by a smooth evolution of the bubble counter for any
time. Yet, these solutions lack universality since they depend on
precise details of the initial conditions.

Nevertheless, there is another class of solutions of the idealized
ST problem: Solution which exhibit a cusp-like singularity after
finite time. A generic ST  bubble always develops a cusp. The cusp
singularity dominates the evolution in its vicinity, and therefore
it is expected to be universal. Thus, our aim is to describe the
local dynamics near cusps of  tip-splitting.

Our central result is, thus, a formula which describes the
evolution of tip splitting in time. It has the form
\begin{eqnarray}
z(s,t) = s^2+u_\phi(t) + \frac{ v_\phi (t)}{w_\phi(t) + i s}. \label{tip-form}
\end{eqnarray}
where $z=x+iy$ is a complex coordinate on the bubble contour,
$-\infty<s<\infty$ parametrize the curve, and $u_\phi(t)$,
$v_\phi(t)$ and $w_\phi(t)$ are functions of the time, $t$. These
functions which will be calculated in what follows, depend on a
single parameter, $\phi$, governing the asymmetric shape of the
evolution. In Fig.~1 we depict contours obtained from
(\ref{tip-form})  which represent snapshots of the tip-splitting
evolution as function of the time.
\begin{figure}
\vspace{-0.0cm}
\includegraphics[width=4cm]{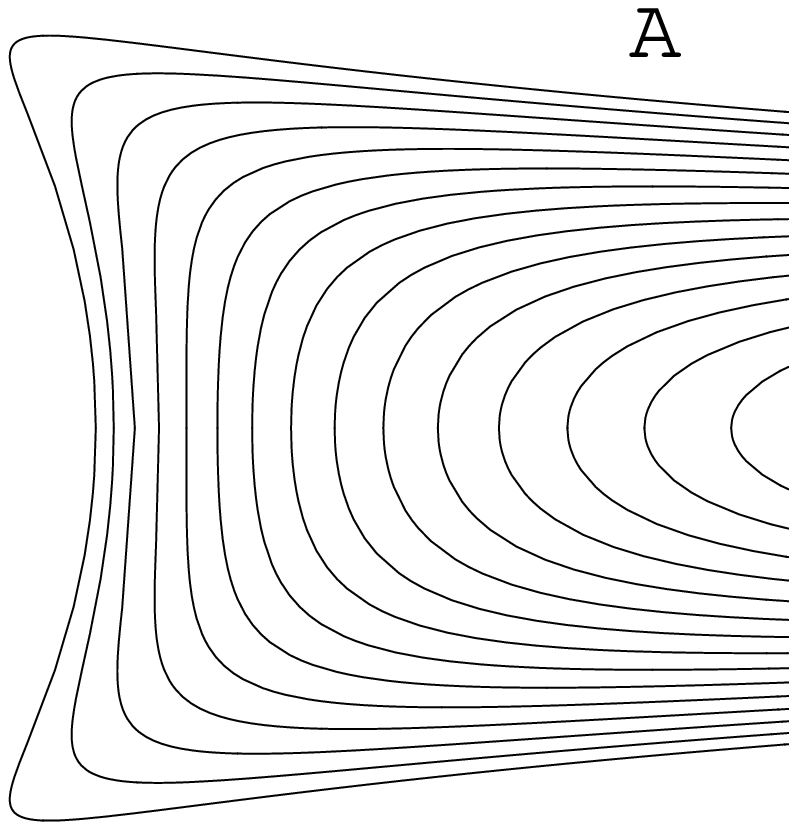}\includegraphics[width=4cm]{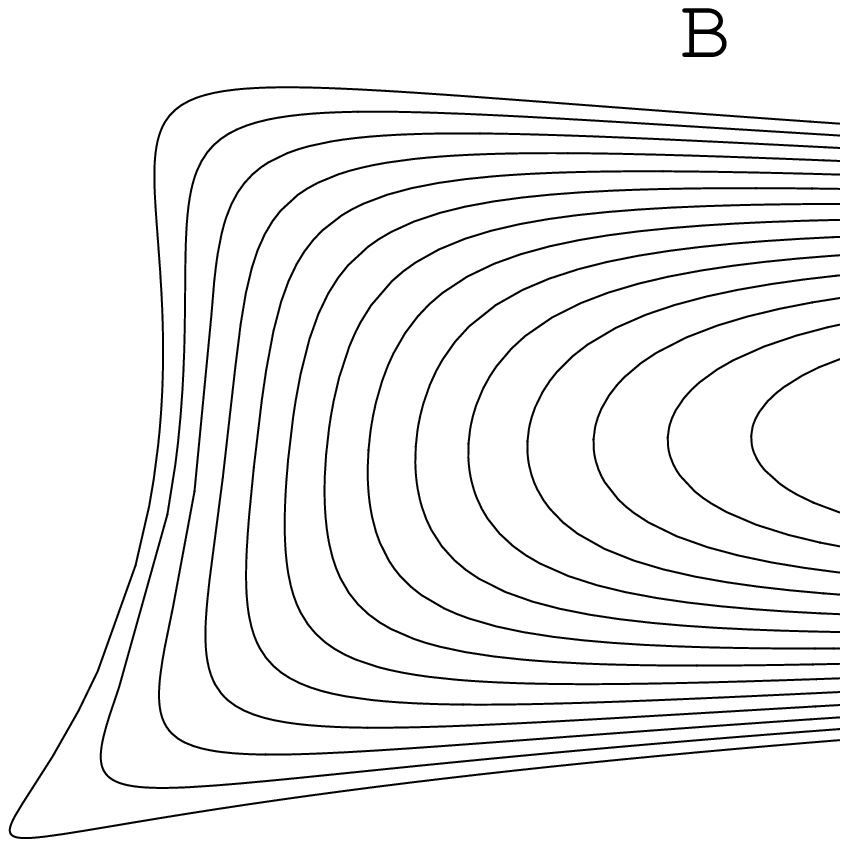}
\caption{The evolution of tip splittings in the Saffman-Taylor problem at zero surface
tension. The contours represent snapshots of the evolution as function of the time.
The evolution given in (\ref{tip-form}), is characterized by one parameter, $\phi$,
which controls the
asymmetry between the two generated fingers. Panel A shows the symmetric
evolution, while panel B represents a typical asymmetric behavior.
\label{cap:tipasymtipsym}}
\vspace{-0.0cm}
\end{figure}

A problem one encounters when trying to describe the tip-splitting
evolution is that a naive extension of the idealized ST dynamics
beyond the cusp singularity is impossible. One may try to compute
the ST dynamics, in the vicinity of a cusp, by introducing some
infinitesimal surface tension. This approach indeed regularizes
the problem  but the resulting equations are nonintegrable, and
the analytical treatment becomes very complicated
\cite{144:Tanveer}. There are some exact steady-state solutions,
with surface tension, but these do not exhibit tip-splitting
\cite{Vasconcelos}.

Alternatively, the cusp singularity may be resolved by a method known as ``dispersive
regularization''\cite{Whitham:1966:NDW}\cite{89:Kodama:Bloch},
which preserves the integrable structure of the problem.
This approach is similar, in spirit, to the construction
of Gurevitch-Pitaevskii solutions for the KdV equation \cite{Gurevich:Pitavsk}.
Its application to the ST problem is discussed extensively
in \footnote{E. Bettelheim et al., in preparation.}. Here we shall
focus on time intervals where these aspects of the regularization are
irrelevant.

We begin by recalling the ST problem. The local velocity of a viscous
fluid in a thin cell is proportional to the gradient of the pressure
$\vec{v}\propto\vec{\nabla}p$, where  $p(z)$ is a harmonic function
due to incompressibility. In the zero surface tension
limit, the pressure may be taken to be equal to zero on
the perimeter of the bubble, while at infinity it diverges
logarithmically. At constant flow rate, the area of the
bubble grows linearly with time, $t$. We set the flow
rate such that the area is $\pi t$. The other parameters, viscosity
and width of the cell, are chosen such that the pressure is equal to
the velocity potential.

An important feature of the ST dynamics, at zero surface tension,
is the conservation of the harmonic moments: $\frac{dt_{k}}{dt}=0$, where
\begin{equation}
t_k=-\frac{1}{\pi k}\int_{\mbox{visc.fluid}}\!\!\!\! \!\!\!\!\!\!\!\!d^2 z z^{-k},
~~k=1,2,\cdots.  \label{harmonic}
\end{equation}
and the integration is over the domain occupied by the viscous fluid.
This property implies that the idealized ST dynamics is integrable, i.e. the
bubble's contour can be determined from the set of harmonic moments, $t_k$'s,
and the area $\pi t$. Our goal is to describe the evolution of this contour
in the vicinity of the tip-splitting.

In order to do, so it is instructive to introduce a potential defined
as $V(z)=t\log(z)+\sum_{k}t_{k}z^{k}$. In particular, it will be convenient
to work with sets of harmonic moments for which
the potential can be resummed  as
\begin{equation}
V(z)=t\log(z)+t_{1}z+\sum_{i=1}^{N}\mu_{i}\log(z-q_{i}).\label{eq:VMiwas}\end{equation}
Thus, the set  $\left\{ t_{k}\right\} _{k=1}^{\infty}$ is replaced
by a  new set of  parameters $\left\{ \mu_{i}\right\} _{i=1}^{N}$
and $\left\{ q_{i}\right\}_{i=1}^{N}$ known as Miwa variables in
the soliton literature\cite{28:Jimbo:Miwa}. In particular $q_{i}$
and $\mu_i$ are, respectively, the location and the weight
(assumed to be real) of the $i$-th Miwa variable.

The potential defined above encodes the shape
of the bubble in a rather complicated way. A function from which
it is easier to extract this information is the
Schwarz function. This function has proved to be a useful tool
for the analysis of the ST problem (see e.g. \cite{Vasconcelos}\cite{EJAM:Howison1}\cite{Schwartz}).
 A Schwarz function, $S(z)$, of a given
contour ${\cal C}$ satisfies the relation $S(z)=\bar{z}$ (where $\bar{z}$ is the complex conjugate of $z$)
for points $z$ which lie on the contour ${\cal C}$. Away from the contour
it is defined via analytic continuation. The simplest example is a
circular contour of area $\pi t$, whose Schwarz function can be
easily deduced to be $S(z)=\frac{t}{z}$.

The Schwarz function is related to the set of harmonic moment by
the contour integral
$t_{k}=\frac{1}{2\pi i}\oint_{{\cal C}}S(z)z^{k}dz.$
It implies that, on the exterior of the bubble, $S(z)$ has the
same singular structure as  $\frac{\partial V(z)}{\partial z}$
(as can be seen by deforming the
contour integral such that it envelopes
the singularities of $S(z)$ on the exterior of the bubble).
Thus the analytic structure of $S(z)$ in the exterior domain, can be
extracted from the potential, $V(z)$.

To reveal the analytic behavior of $S(z)$ in the interior of the bubble
one may employ an important property of the Schwarz function known as the
unitarity condition,
\begin{equation}
\bar{S}(S(z))=z, \label{unitarity}
\end{equation}
where the complex conjugation of a function, $\bar{f}(z)$, is defined
as $\overline{f}(z)\equiv\overline{f(\overline{z})}$. The equation above trivially
holds for $z\in{\cal C}$ and by analytic continuation all over the complex plane.
$S(z)$ maps points from
the exterior of the bubble (where the analytic structure is known)
to points on the interior domain. Thus all
singularities of $S(z)$ can be identified, and hence its precise
analytic form, as function of the Miwa variables
and the time.

We proceed by assuming that the Schwarz function is an algebraic
function, i.e.~a function defined on some algebraic Riemann
surface. Let us explain what is an algebraic Riemann surface, and
then describe how to define a Schwarz function on this surface. Consider a polynomial
equation of the form $P(z,f(z))=0,$ where $P$ is a polynomial in two variables.
Solving this equation we obtain for each $z$ a set of solutions $f(z)$.
In general $f(z)$ will have branch cuts. A simple example of such
a polynomial is $P(z,f)=f^{2}-\Pi_{i=1}^{4}(z-\lambda_{i})$, where
we obtain $f(z)=\sqrt{\Pi_{i=1}^{4}(z-\lambda_{i})}$. In general
there will be $n$ solutions of $f$ for each $z$. Thus one may introduce
$n$ copies of the complex plane, where each copy is associated with
a well defined value of $f(z)$. Clearly, on each copy of the complex
plane $f(z)$ is discontinuous along the branch cuts. To define it as a
continuous function we may paste together the various sheets of the complex plane along
the branch cuts. Then, together with a choice of local coordinate
systems around each point one obtains a Riemann surface composed of
$n$ sheets.

Since we assume that the Schwarz function may be defined on such
a Riemann surface, one should assign for each point on the
Riemann surface, a value of Schwarz function, $S$. In order to specify
such a point we must indicate the sheet-index $i$
(where $1\leq i\leq n$), and a complex number $z$. The index, $i$,
will specify on which copy of the complex plane the point lies,
and a complex number, $z$, will specify the coordinate on
that copy. There will be one copy of the complex plane, which will
be termed as ``the physical sheet'', on which the bubble lies.
On this Riemann sheet $S(z)$ will be equal to $\bar{z}$ on the perimeter
of the bubble. All other ``unphysical'' Riemann sheets will be considered
 as parts of the  ``interior domain'' since there are  no branch
cuts on the exterior domain (where the pressure is harmonic).

Let us now consider a situation where the potential has the form
(\ref{eq:VMiwas}), and show that the number of unphysical sheets
is the number of Miwa variables. On the exterior of the bubble
$S(z)$ has the following singular structure determined by
$\frac{\partial V}{\partial z}$:\[
S(z)\sim\frac{t}{z}+t_{1}-\sum_{i=1}^{N}\frac{\mu_{i}}{z-q_{i}}.\]
For the point $q_{i}$, unitarity (\ref{unitarity}) implies that
$\bar{S}(S(q_{i}))=\bar{S}(\infty)=q_{i}.$ Thus a point at
infinity should be mapped to the point $\bar{q}_{i}$. This point
should be located on one of the unphysical sheets since on the
physical sheet $S(\infty)=t_{1}$. Conversely, the value of the
Schwarz function at infinity on an unphysical sheet correspond to
a particular Miwa variable. From here on we consider the case of
two Miwa variables, which is the minimal model exhibiting a tip
splitting scenario. The corresponding Riemann surface is thus
composed of three sheets.

For actual calculation of the contour's evolution it will be convenient to
use a conformal mapping which maps the exterior of some ``source domain'' (in $\zeta$-plane)
to the exterior of the physical bubble, the ``target domain'' (in $z$-space).
The source domain is usually taken to be the unit circle, however, we found it
more convenient to use a bubble with a cusp.
The advantage in using this source domain is that we can choose the
mapping to be non-trivial only  around the tip, while everywhere
else it would be approximately proportional to the identity map.

We will study the evolution near the tip in the
case where the weight of one of the Miwa variables,
is small compared to the other. This suggests choosing the source
domain to be the one Miwa-variable bubble at the point where
a cusp is formed. The Riemann surface in this case
is composed of two sheets as described above.
If we assume this surface to be of genus zero, then by the Riemann-Hurwitz
theorem, the number of branch points is two. Thus the Riemann surface
is given by the polynomial equation  $R^{2}=(\zeta-\lambda_{1})(\zeta-\lambda_{2})$.
A function defined on this surface (the Schwarz function in particular)
contains a branch  cut which extends from point $\lambda_{1}$ to the point $\lambda_{2}$.
An example of such a function is $R(\zeta)$.

The 1-Miwa bubble is characterized by four parameters $t_{1},
t,q$  and $\mu$, since the location of the branch points depends on these
parameters, one can calculate the area of the bubble and its first
harmonic moment, $t_{1}$, from the location of the branch points
$\lambda_{1}$ and $\lambda_{2}.$ We will fix $\mu=-q=1$ and
treat $\lambda_{1}$ and $\lambda_{2}$ as the parameters which describe
the bubble. If we also demand that the bubble is at the moment
where the cusp forms, we can characterize the bubble by a single parameter,
say $\lambda_{1}$. We assume that $\lambda_{1}< \lambda_{2} $ and that $\lambda_{1}$ is of
order $\delta,$ where $\delta\ll 1$.
This assumption implies that the global bubble shape is dominated by the cusp.
Then the bubble area is of order $\delta^{3}$ while the
first harmonic moment, $t_{1},$ is $1+O(\delta^{2})$. The Schwarz function, $\sigma(\zeta)$,
of such a bubble is:
\[
2\sigma(\zeta)=\left(t_{1}-1\right)+\frac{t-1}{\zeta-\bar{t}_{1}}-
\frac{1}{\zeta+1}+\frac{(t_{1}+1)(\zeta-\lambda_1)R(\zeta)}{(\zeta-t_1)(\zeta+1)},\]
 as can be ascertained by examining this function's singular structure
and that it satisfies the unitarity condition (\ref{unitarity}).
The fact that this  function describes a bubble with a cusp can be
checked by considering the behavior of the solution of $\sigma(\zeta)=\bar{\zeta}$
near $\lambda_{1}$ (where the cusp is located).

Up till now we have characterized the ``source domain''.
We would like, now, to specify the physical bubble, or the ``target
domain'' associated with two Miwa variables.
For this purpose we must give the mapping
between the source and target domains. This mapping is taken to be:
\begin{equation}
z(\zeta)=c_{1}\left(\zeta+\frac{\alpha}{2}
\frac{R(\theta)-R(\zeta)}{\zeta-\theta}+\beta R(\zeta)\right)+c_{2}, \label{mapping}
\end{equation}
where $c_1$, $c_2$, $\alpha$, $\beta$, and $\theta$ are parameters of the mapping and
$R(\zeta)\equiv\sqrt{\zeta-\lambda_1}\sqrt{\zeta-\lambda_2}$.
This mapping can be considered as a mapping from the 1-Miwa Riemann surface
to the 2-Miwa  Riemann surface.  The mapping has singularities at the infinities
on each of the sheets of the Riemann surface associated with the source
domain, which are mapped to infinites on different sheets on the target
domain, as well as at the point $\theta$ on the unphysical sheet. This
point, $\theta,$ is mapped to an infinity on a third sheet
of the target Riemann surface. Thus (\ref{mapping}) indeed
maps a two-sheeted Riemann surface, which is associated with a 1-Miwa
bubble, to a three sheeted Riemann surface.

That, indeed, the target Riemann surface is associated
with a 2-Miwa variable bubble, can be deduced from the
singular structure of the Schwarz function
of the target domain. The latter is given by
$S(z)=\bar{z}\left(\sigma\left(\zeta\left(z\right)\right)\right)$, where
$\zeta(z)$ is the inverse map of $z(\zeta)$, and $\sigma(\zeta)$ is the Schwarz function
of the source domain.

Having $S(z)$ one may extract all constants of motion
(Miwa variables, $q_1$, and $q_2$; Miwa weights, $\mu_1$ and $\mu_2$; and $t_1$),
and the area $t$, from its singular structure.
These will be expressed as functions of the parameters of the mapping (\ref{mapping})
and $\lambda_1$. Solving these relations one may express the parameters of the mapping
as functions of the time and thus obtain the evolution of the contour.
We take $\mu_1=-q_{1}=1$ by fixing $c_{1}$ and
$c_{2}$. Thus the parameters whose time evolution is to be determined are $\alpha$, $\beta$,
$\theta$ and  $\lambda_{1}$.

To obtain reasonably simple equations, we expand
all quantities in orders of $\delta$ and take the leading order.
Let us assume that for some initial moment, $t^{(0)}$,
around the formation of the tip, $\lambda_{1}$ assumes the value $\lambda$,
to leading order in $\delta$. We now make the following scaling ansatz:
$\alpha\sim\delta^{4}$, $\beta\sim\delta^{3}$, $\nu\equiv\lambda_{1}-\theta\sim\delta^{3}$ and
$\delta\lambda\equiv\lambda_{1}-\lambda\sim\delta^{3}$. Then the
equations we obtain, to leading order in $\delta$, for the constants
of motion $q_{2}-\lambda$ and $\mu_{2}$ are:
\begin{eqnarray}
q_{2}-\lambda=\beta+\delta\lambda-\bar{\nu}-\frac{\alpha\sqrt{-\lambda}}{\sqrt{\nu}+\sqrt{\bar{\nu}}}~~~~~~~ \label{first}\\
\mu_{2}=-2\bar{\alpha}\sqrt{-\lambda}\sqrt{\nu}\left(1-\frac{\alpha\sqrt{-\lambda}}{2\sqrt{\bar{\nu}}\left(\sqrt{\nu}+\sqrt{\bar{\nu}}\right)}\right)
\end{eqnarray}

Let us now define  $T_{1}=t_1-\left( 1+3\lambda^{2}-7\lambda^{3}+\frac{33}{2}
\lambda^{4}\right)$ (the difference between the first harmonic moment
of the target bubble and a source bubble with $\lambda_1=\lambda$ to order $O(\delta^5)$),
and similarly $T=t-(-4\lambda^{3}+18\lambda^{4}-63\lambda^{5})$.
With these definitions, analysis of the singularities of $S(z)$ yields:
\begin{eqnarray}
T=2\Re(\alpha)\lambda+4\beta\lambda^{2}-12\lambda^{2}\delta\lambda,\\
T_{1}=\Re(\alpha)-2\beta\lambda+6\lambda\delta\lambda.  \label{last}
\end{eqnarray}

The solution of Eqs.~(\ref{first}-\ref{last}) give the parameters
$\alpha,$ $\beta,$ $\nu$  and $\delta\lambda$ in terms of $q_{2},$ $\mu_{2}$,
$T$ and $T_{1}$, which define, in turn, the conformal mapping of the contour on the target space
as function of the time. To write down the solution of these equations it will be convenient
to define a shifted rescaled time $\delta t\equiv-\frac{T+2T_{1}\lambda}{4\mu_{2}^{3/4}\sqrt{-\lambda}}$
and  introduce two functions $\xi$ and $\eta$ which satisfy the nonlinear equations:
\[
\left(\frac{\sqrt{2}\delta t}{\sqrt{\xi}}+\frac{\delta t^{2}}{2\xi^{2}}\right)
\left(\eta+\xi\right)=1;~~\eta=\left(\frac{\phi}{\sqrt{\xi}-\frac{\delta t}{\sqrt{2}\xi}} \right)^2,\]
where $\phi=\Im (\frac{q_2}{\sqrt{\mu_2}}))$ is the asymmetry parameter. The solution of these equations
gives $\eta$ and $\xi$ as functions of the rescaled time, $\delta t$, and $\phi$.
With the help of these functions we may write the solution of Eqs.~(\ref{first}-\ref{last}) as:
\begin{eqnarray*}
\alpha=\frac{\mu_{2}^{3/4}\delta t}{\sqrt{-\lambda}}\left(-1+i\sqrt{\frac{\eta}{\xi}}\right),~~~~~~~~~~~~~~~~~
~~~~~~~~~~~~\\
4\beta =3\Re(q_{2})-3\lambda-\frac{3}{2}(\eta-\xi)-\frac{3\delta t}{\sqrt{2\xi}}-\frac{T_{1}}{4\lambda}+
\frac{T}{8\lambda^{2}}, \\
\nu=\frac{\xi-\eta}{2}+i\sqrt{\xi\eta},~~~~~~~~~~~~~~~~~~~~~~~~~~~~~~~~~~~~~~~~ \\
4\delta \lambda =3\Re(q_{2})-3\lambda-\frac{3}{2}(\eta-\xi)-\frac{3\delta t}{\sqrt{2\xi}}+\frac{T_{1}}{4\lambda}-
\frac{T}{8\lambda^{2}}. \\
\end{eqnarray*}

Given this time dependence of the parameters of the conformal
mapping we are in a position to describe the contour dynamics in
the vicinity of the tip-splitting. For this purpose it is
sufficient to focus on the image of the source domain around the
cusp. The shape of the source domain near the cusp is given by the
universal form $y=A\delta x^{3/2}$, were $\delta x\equiv
x-\lambda_1$, and $A$ is some constant. Close enough to the cusp
$y\ll\delta x,$ and therefore we may assume that $y\simeq0$ (this
assumption can be proved to be consistent with the expansion in
the parameter $\delta$ performed above). Thus one has to find the
image of the ray $x>\lambda_{1}$ under the mapping $z(\zeta)$ to
leading order in $\delta$. The result is given by
Eq.~(\ref{tip-form}), where the functions $u_\phi(t)$,
$v_\phi(t)$, and $w_\phi(t)$ are:
\begin{eqnarray*}
u_\phi(t)= \frac{\xi-\eta}{2}- \frac{\delta t}{\sqrt{2 \xi}}, \\
v_\phi(t)=\delta t\left( 1-i\sqrt{\frac{\eta}{\xi}}\right), \\
w_\phi(t)= \frac{\xi-\eta}{2} + i \sqrt{\eta \xi}.
\end{eqnarray*}
The above equations together with (\ref{tip-form}) describe the
evolution of a tip-splitting of the  ST problem at zero surface
tension (Fig.~1). The form of the tip-splitting depends on a
single parameter, $\phi$, which controls the asymmetric shape of
the evolution. Since our derivation of the tip-splitting formula
makes use only of local properties near the tip, it is suggestive
that this evolution, for short times, is universal. Namely, the
tip splitting evolution is characterized by one parameter, $\phi$,
independent of the shape of the bubble on large scales. The
shortcomings of our analysis is that it does not include the
influence of surface tension. Therefore it breaks down after a
short time due to the formation of cusps. The method of dispersive
regularization can be employed again to continue the evolution.
Thus, in comparing our theoretical prediction with experiment, one
can hope for agreement only within a limited time interval near
the initial stage of the tip-splitting.

We thank Paul Wiegmann and Anton Zabrodin for useful discussions.
This research has been supported in part by the Israel Science
Foundation (ISF), and by the German-Israel Foundation (GIF).

\vspace{-0.0cm}
\bibliography{mybib}

\end{document}